\documentclass[conference]{IEEEtran}
\IEEEoverridecommandlockouts

\usepackage{cite}
\usepackage{amsmath,amssymb,amsfonts,amsthm}
\usepackage{graphicx}
\usepackage{textcomp}
\usepackage{xcolor}
\def\BibTeX{{\rm B\kern-.05em{\sc i\kern-.025em b}\kern-.08em
    T\kern-.1667em\lower.7ex\hbox{E}\kern-.125emX}}

\usepackage{cleveref}
\usepackage{tabularray}
\usepackage{booktabs}
\usepackage{multirow}
\usepackage{makecell}

\usepackage{tabularray}
\usepackage{booktabs}
\usepackage{multirow}
\usepackage{makecell}

\usepackage{algorithm}
\usepackage[noend]{algpseudocode}
\makeatletter
\def\BState{\State\hskip-\ALG@thistlm}
\makeatother
\newcommand{\R}{\mathbb{R}}

\newtheorem{lemma}{Lemma}

\newtheorem{proposition}{Proposition}
\newtheorem{definition}{Definition}

\usepackage{matlab-prettifier}
\usepackage{color}
\usepackage{mathrsfs}
\usepackage[shortlabels]{enumitem}
\usepackage{subcaption}

\begin{document}

\title{\LARGE \bf PACT: A Contract-Theoretic Framework for Pricing Agentic AI Services Powered by Large Language Models 
}

\author{
Ya-Ting Yang  and  Quanyan Zhu\\
Department of Electrical and Computer Engineering, 
Tandon School of Engineering, \\ New York University, 
Brooklyn, NY, USA; 
\texttt{\{yy4348, qz494\}@nyu.edu}
\thanks{This work has been submitted to the IEEE for possible publication.
Copyright may be transferred without notice, after which this version may no longer be accessible.}
}

\maketitle

\begin{abstract}
Agentic AI, often powered by large language models (LLMs), is becoming increasingly popular and adopted to support autonomous reasoning, decision-making, and task execution across various domains. While agentic AI holds great promise, its deployment as services for easy access raises critical challenges in pricing, due to high infrastructure and computation costs, multi-dimensional and task-dependent Quality of Service (QoS), and growing concerns around liability in high-stakes applications. In this work, we propose PACT, a Pricing framework for cloud-based Agentic AI services through a Contract-Theoretic approach, which models QoS along both objective (e.g., response time) and subjective (e.g., user satisfaction) dimensions. PACT accounts for computational, infrastructure, and potential liability costs for the service provider, while ensuring incentive compatibility and individual rationality for the user under information asymmetry. Through contract-based selection, users receive tailored service offerings aligned with their needs. Numerical evaluations demonstrate that PACT improves QoS alignment between users and providers and offers a scalable, liable approach to pricing agentic AI services in the future.
\end{abstract}

\begin{IEEEkeywords}
Pricing, Large Language Models, Agentic AI, Contract Theory
\end{IEEEkeywords}

\section{Introduction}
Agentic AI \cite{acharya2025agentic}, often powered by large language models (LLMs), is gaining significant attraction in both industry and academia due to its wide range of applications that enable autonomous decision-making, streamline complex workflows, and assist with tasks such as reasoning, planning, and real-time adaptation on behave of users or organizations across various domains, including customer service, cybersecurity, healthcare, and business operations. Thanks to recent advancements in wireless communication and cloud computing \cite{10811953}, along with inspiration from \cite{ahmaditeshnizi2023optimus}, these agents can be instantiated as cloud-based services \cite{liagkou2024cost}, embedded service applications, or distributed systems that operate across digital and physical environments \cite{yin2024llm}. They process language-based input and generate human-readable output, facilitating easier access to AI-based techniques. For example, retail shop owners can use such a service to retrieve relevant information and receive recommendations for profit maximization, without the need to formulate or solve the optimization problem themselves.

However, service providers may lack sufficient incentives for continuous service development, quality assurance, and technical maintenance due to the substantial costs associated with computation, hardware infrastructure, and energy consumption required to deploy and operate LLM-based services \cite{bergemann2025economics}. As a result, agentic AI services needs to be offered through an efficient pricing model. Such a pricing scheme not only helps recover operational costs but also allows for differentiated service levels based on factors such as response time or reasoning complexity, enabling providers to align service quality with different user needs. In addition, pricing helps discourage overuse, encourages responsible consumption, and supports accountability, particularly in high-stakes applications \cite{gallagher2024assessing}.

\begin{figure}
    \centering
    \includegraphics[width=0.99\linewidth]{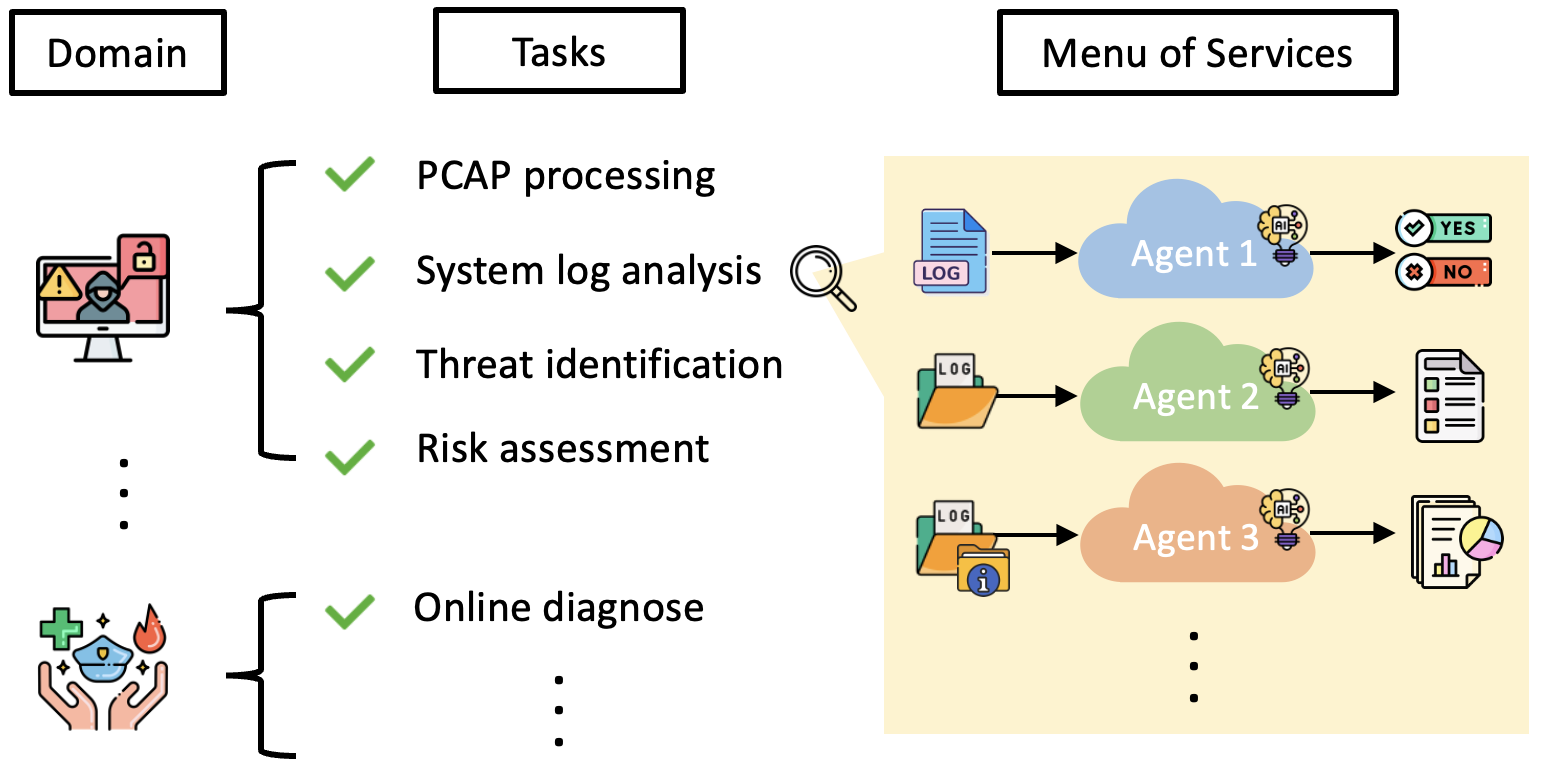}
    \caption{An illustration of cloud-based agentic AI services for tasks in different domains.}
    \label{fig:menu}
\vspace{-3mm}
\end{figure}

A key challenge in pricing LLM-based services lies in their inherently multi-dimensional Quality of Service (QoS). Unlike traditional cloud-based services, where performance metrics such as latency or accuracy can be measured, the value of LLM outcomes depends on a complex interplay of attributes. These include not only response time but also cognitive depth (e.g., Chain-of-Thought reasoning \cite{NEURIPS2022_9d560961}), factual accuracy, and alignment with user preferences \cite{wang2024understanding}. The importance of each attribute may vary across tasks, making QoS both multi-dimensional and context-dependent. This complexity introduces another challenge: estimating user satisfaction with service outcomes. Since LLM performance is frequently subjective and must be inferred from indirect signals such as downstream task success, explicit user feedback, or behavioral cues like query reformulation \cite{liu2023agentbench}, pricing must be informed by satisfaction models that may leverage machine learning to infer user satisfaction from these signals \cite{lin2024interpretable}.

In addition to technical considerations, regulatory and ethical concerns can also influence LLM pricing. LLM-based agentic AIs are increasingly deployed in high-stakes domains such as cybersecurity, healthcare, and the law sectors, where hallucinations or misalignment can have serious consequences \cite{liu2024exploring}. This raises critical questions: Should service providers be held liable for harms caused by generated content? Should users be charged more for auditable or higher-assurance outputs? These considerations highlight that LLM pricing is not solely a market mechanism but rather a socio-technical issue, demanding alignment with legal, ethical, and societal expectations.

To address these challenges, we propose the PACT framework (Pricing scheme for cloud-based Agentic AI services through a Contract-Theoretic perspective). In this framework, the task-dependent QoS is defined primarily by two dimensions: response time from the cloud and estimated user satisfaction with the service outcome. The QoS model can be extended to accommodate additional dimensions depending on the task. From the service provider's perspective, we account not only for computational and infrastructure costs but also for potential liability costs associated with future regulatory and ethical obligations. Furthermore, the contract-based incentive mechanism \cite{zhang2015contract,chen2023qos} ensures incentive compatibility, encouraging users to select QoS levels that align with their true preferences and task requirements. 

To this end, our contribution can be summarized as follows. We propose PACT, a contract-theoretic pricing framework tailored for cloud-based agentic AI services, which models task-dependent multi-dimensional QoS for agentic AI as well as integrates computational, infrastructure, and liability costs into the pricing structure, accounting for future regulatory and ethical risks in the high-stakes domains. The contract-based method ensures incentive compatibility and individual rationality, allowing heterogeneous users to select services aligned with their needs while maximizing the provider’s utility under asymmetric information. Numerical experiments validate the framework’s effectiveness in aligning QoS with user preferences while balancing the service provider's utility and liability.

\section{System Model}
We focus on designing a pricing scheme for cloud-based LLM services (AI agents). In the proposed framework, each user acquires LLM services from the service provider (SP) to meet their specific task-oriented needs.For example, as illustrated in Fig. \ref{fig:menu}, cybersecurity users (defenders) may subscribe to cloud-based services designed for tasks such as system log analysis, PCAP data processing, proactive threat identification, risk assessment, and defense strategy recommendations. The set of tasks offered by the SP is denoted as $\Phi$, where each task is represented by $\phi \in \Phi$.
However, even for the same task $\phi$, such as system log analysis, different users may have varying requirements for the quality of service (QoS), which in turn leads to different payments \cite{chen2023qos}. For example, a security analyst responsible for protecting critical infrastructure in an enterprise network may require better QoS compared to a user monitoring access to personal or non-sensitive data.
In this context, let $\mathcal{S}^\phi=\{s^\phi_1, \cdots, s^\phi_k, \cdots, s^\phi_{K}\}$ represent the set of possible $K$ services that the SP can offer for task $\phi$,  where each service corresponds to a specific QoS level. 

Specifically, in this study, the quality is characterized by the response time and the user's satisfaction with the response \cite{MicrosoftLLM2023}. For instance, the latency in the LLM service's response can impact how quickly a user can react to emerging threats, with some users requiring faster responses than others. Additionally, users may demand different levels of detail in the results: some may be satisfied with a simple yes/no answer, while others may need a detailed explanation or a chain-of-thought (CoT) reasoning behind the response. In the following sections, we focus on pricing different QoS levels for a specific task $\phi$ and will therefore omit the superscript $\phi$ for simplicity.

\subsection{Latency of the LLM Service}
The latency in a cloud-based LLM service's response typically consists of transmission time, tokenization and detokenization time, and model inference time. Let $D_k^{\text{in}}$ and $D_k^{\text{out}}$ (KB) denote the input and response data size for service $s_k$, respectively, and let $r$ (bps) be the communication rate between the user and the cloud. The transmission time can then be written as:
\begin{equation}
    T_k^{\text{tran}}= \frac{8000(D_k^{\text{in}} + D_k^{\text{out}})}{r}.
\end{equation}
Then, tokenization time generally depends on the number of characters or words in the input text and the tokenization method (e.g., word-level, subword-level, character-level, Byte Pair Encoding (BPE), etc.) being used. Similarly, detokenization time depends on the number of tokens and the detokenization method. As the number of characters is typically proportional to the data size (1 or 2 bytes per text character, depending on ASCII or Unicode encoding), we have
\begin{equation}
    T_k^{\text{tok}}= \alpha_{t}D_k^{\text{in}}+ \alpha_{dt} D_k^{\text{out}},
\end{equation} where $\alpha_{t}, \alpha_{dt}$ are constant parameters that depend on the complexity of the tokenization and detokenization algorithms.

If the number of parameters of the LLM model used for service $s_k$ is denoted by $\beta_k$, according to OpenAI \cite{kaplan2020scaling}, the floating point of operations (FLOPs) per token for the transformer-based model is: $2\beta_k + 2n^{\text{layer}}_k n^{\text{ctx}}_k n^{\text{attn}}_k$, where $n^{\text{layer}}_k$, $n^{\text{ctx}}_k$, and $n^{\text{attn}}_k$ are the number of layers, the maximum number of input tokens, and the dimension of the attention output of the model used for providing $s_k$, respectively. Let $g_t: \R_{\ge 0} \mapsto \R_{\ge 0}$ represent the mapping between the input data size and the number of input tokens after tokenization, and $g_{dt}: \R_{\ge 0} \mapsto \R_{\ge 0}$ for the mapping between the number of output tokens and the output data size after detokenization. Denote the computation capability of the cloud server as $\gamma_k$ giga floating point operations per second (GFLOPS), the inference time can then be approximated as:
\begin{equation}
    T_k^{\text{inf}} = \frac{[g_t(D_k^{\text{in}})+g_{dt}(D_k^{\text{out}})]\cdot(2\beta_k + 2n^{\text{layer}}_k n^{\text{ctx}}_k n^{\text{attn}}_k)}{\gamma_k}.
\end{equation}
To this end, the total latency for service $s_k$ is 
\begin{equation}
    T_k = T_k^{\text{tran}} + T_k^{\text{tok}} + T_k^{\text{inf}}.
\end{equation}

\subsection{User Satisfaction with the LLM Service}
For many real-world tasks, such as log analysis and strategy recommendation in cybersecurity, there is rarely a single ``correct'' response. In these cases, the accuracy of the LLM service should be measured by how well it aligns with user preferences and its ability to fulfill the user's intent. Let the users' expected response be denoted as $e$, The user’s satisfaction, $A^k \in [0, 1]$, with service $k$ is then captured by the probability that the output response $D^{\text{out}}_k$ meets $e$. That is, 
\begin{equation}
    A_k=\textbf{Pr}(D^{\text{out}}_k \text{ meets } e).
\end{equation} Note that $A_k$ can be estimated through user surveys \cite{wang2024understanding}, completion percentage of downstream tasks \cite{liu2023agentbench}, or learned through machine learning methods \cite{lin2024interpretable}.

\subsection{Quality of the LLM Service}
With latency $T_k$ and user satisfaction $A_k$, an option for quantifying the quality of service $s_k$ is:
\begin{equation}
    q_k = \delta A_k + (1- \delta)(1- T_k),
\end{equation} 
where the parameter $\delta \in [0, 1]$ reflects the task-specific importance of user satisfaction relative to latency. A higher value of $\delta$ places greater emphasis on user satisfaction, which is suitable for subjective tasks such as report summarization or strategy recommendation. Conversely, a lower $\delta$ prioritizes latency, which is critical for real-time tasks such as intrusion detection or threat alerting. Since each service $s_k$ is associated with a corresponding QoS level $q_k$, we denote the set of all possible QoS levels provided by the SP (for task $\phi$) as $\mathcal{Q}=\{q_1, \cdots, q_k, \cdots, q_{K}\}$ which will be used in the subsequent analysis.

\subsection{Cost of the LLM Service}
The cost for the SP to provide an LLM service 
$s_k$ with QoS level $q_k$ typically consists of computational (energy) consumption as well as hardware infrastructure and model deployment cost. It can be represented by the cost function $C:\mathcal{Q} \mapsto \mathbb{R}$. Specifically, $C(q_k)$ can be expressed as:
\begin{equation}
    C(q_k) = c^{\text{tok}}(q_k) + c^{\text{h}}(q_k) + c^{\text{m}}(q_k)+c^{\text{l}}(q_k),
\label{eq:c_q}
\end{equation} where $c^{\text{tok}}:\mathcal{Q} \mapsto \mathbb{R}$ is the computational cost associated with the number of input and output tokens for providing $q_k$. $c^{\text{h}}:\mathcal{Q} \mapsto \mathbb{R}$ is the hardware infrastructure such as the cloud server equipped with different types of GPU/TPU costs, $c^{\text{m}}:\mathcal{Q} \mapsto \mathbb{R}$ represents other LLM model-related costs, and $c^{\text{l}}:\mathcal{Q} \mapsto \mathbb{R}$ indicates the liability costs associated with the service. Specifically, with $\beta_k$ being the number of parameters of the model used for providing service $s_k$ with QoS level $q_k$, the floating point of operations (FLOPs) per token for the transformer-based model is again: $2\beta_k + 2n^{\text{layer}}_k n^{\text{ctx}}_k n^{\text{attn}}_k$. In this case, when the number of input and output tokens is $[g_t(D_k^{\text{in}})+g_{dt}(D_k^{\text{out}})]$, the cost $c^{\text{tok}}(q_k)$ can be expressed as:
\begin{equation*}
    c^{\text{tok}}(q_k) = c\cdot \big[g_t(D_k^{\text{in}})+g_{dt}(D_k^{\text{out}})\big] \cdot \big( 2\beta_k + 2n^{\text{layer}}_k n^{\text{ctx}}_k n^{\text{attn}}_k\big),
\end{equation*} where $c \in \R_{+}$ denotes the monetary computational (energy) cost per floating-point operation (FLOP). The hardware and model-related costs, $c^{\text{h}}(q_k)$ and $c^{\text{m}}(q_k)$, can be treated as fixed subscription-based fees. Here, $c^{\text{l}}(q_k)$ aims to capture the potential liability for being held accountable for harmful outcomes (e.g., hallucination, misdiagnoses, misinformation, bias, or enabling malicious use). Assessing liability and its cost requires further investigation \cite{gabison2025inherent}, as it depends on factors such as the task's inherent risk, the explainability of the service outcomes, and the legal and regulatory frameworks.

\section{Contract Design for Service Pricing}

\subsection{The User's Type and Utility}
In this work, the user's preference over QoS is characterized by its type, which represents the importance level of the user's task depending on the LLM QoS. For instance, a security analyst responsible
for protecting critical infrastructure in an enterprise network may willing to pay more for better QoS compared to a user monitoring access to non-sensitive data. Here, user types belong to a discrete and finite space.

\begin{definition}[User Types]
Consider a task with service set $\mathcal{S}$, where $|\mathcal{S}|=K$, the users' preferences are sorted in ascending order and classified into a set of $K$ types, denoted as $\Theta=\{\theta_1$, $\cdots$, $\theta_k$, $\cdots$, $\theta_K$\}, and follow the relationship
\begin{equation}
    \theta_1 < \cdots < \theta_k < \cdots < \theta_K , \quad i \in \{2, \cdots, K-1\}.
\end{equation}
\label{def:types}
\vspace{-5mm}
\end{definition}
In this context, a higher $\theta_k$ implies more willingness to pay for higher QoS. Here, the contract designed for type $\theta_k$ user is denoted as $(q_k, p_k)$.

Then, the utility function of a user is the
(monetary) valuation of the QoS minus the payment paid to the SP. Hence, a user with type $\theta_k$ signing contract $(q_k, p_k)$ has the utility $u_k: \mathcal{Q} \times \mathbb{R}_{\ge 0} \mapsto \mathbb{R}$, with
\begin{equation}
    u_k(q_k, p_k) = \theta_k v(q_k) - p_k,
\label{eq:u_i}
\end{equation} where $v:\mathcal{Q} \mapsto \mathbb{R}$ is the (monetary) evaluation function for the QoS, which is a strictly increasing concave function of $q$, where $v(0) = 0, v'(q) > 0$, and $v''(q) < 0$ for all $q \in \mathcal{Q}$. 

\subsection{The Service Provider's Profit}
The SP's utility gained from contract $(q_k, p_k)$ consists of payment $p_k$ made from the user minus the cost $C(q_k)$ of providing the LLM service $s_k: u_{SP}(q_k, p_k) = p_k-C(q_k)$.
Note that the SP may not have direct knowledge of each user’s type, but it can infer the overall distribution of types based on statistical insights from the LLM or agentic AI service market. Hence, instead of knowing the exact type of any given user, we assume the SP has access to the joint probability mass function (PMF) of the user type distribution, denoted as $P(\theta_k)$. Then, the total profit of the SP obtained from the user is:
\begin{equation}
    U_{SP} = \sum_{k=1}^{K}P(\theta_k)u_{SP}(q_k, p_k).
    \label{eq:U_sp}
\end{equation}

\subsection{Optimal Contract}
A contract is feasible if and only if each user selects the contract item specifically designed for them. In particular, feasibility is ensured when the Individual Rationality (IR) and Incentive Compatibility (IC) constraints defined in the following are satisfied simultaneously.

\begin{definition} (Individual Rationality (IR))
Consider a set of user types $\Theta$ defined in Definition \ref{def:types}, a menu of contracts, denoted as $\Omega(\Theta)=\{(q_k, p_k), \theta_k \in \Theta\}$, is individually rational if, for the user of type $\theta_k$, selecting the contract $(q_k, p_k)$ designed for its type has non-negative utility, i.e.,
\begin{equation}
    u_{k}(q_k, p_k) \geq 0, \quad \forall \theta_k \in \Theta.
\label{eq:cons_IR}
\end{equation}
\label{def:cons_IR}
\vspace{-5mm}
\end{definition}

\begin{definition} (Incentive Compatibility (IC)) 
Consider a set of user types $\Theta$ defined in Definition \ref{def:types}, a menu of contracts, denoted as $\Omega(\Theta)=\{(q_k, p_k), \theta_k \in \Theta\}$, is incentive compatible if, for the user of type $\theta_k$, selecting the contract $(q_k, p_k)$ designed for its type maximizes its utility, i.e.,
\begin{equation}
    u_{k}(q_k, p_k) \geq u_{k}(q_j, p_j), \quad \forall \theta_{k}, \theta_{j} \in \Theta^2.
\label{eq:cons_IC}
\end{equation}
\label{def:cons_IC}
\vspace{-5mm}
\end{definition}
The goal of the SP is to determine the pricing scheme $p_{k}$ for the corresponding service $s_{k}$ that has quality $q_k$ jointly that yields it the best return. To this end, the SP is to solve the following constrained optimization problem:
\begin{equation}
    \begin{aligned}
        \max_{\Omega(\Theta)} \quad & \sum_{k=1}^{K}P(\theta_k)u_{SP}(q_k, p_k)\\
        \quad \text{s.t.} \quad & u_{k}(q_k, p_k) \geq 0, \ \forall \theta_k \in \Theta \ \text{(IR)},\\
        & u_{k}(q_k, p_k) \geq u_{k}(q_j, p_j), \ \forall \theta_{k}, \theta_{j} \in \Theta^2, k \neq j \ \text{(IC)}.
    \end{aligned}
\label{prob:opt_contract}
\end{equation}

One challenge in solving the problem above is the presence of $K$ IR and $K \times (K - 1)$ IC constraints. To solve the contract design problem efficiently, we may need to reduce the number of constraints through standard contract analysis \cite{zhang2015contract}.

\begin{lemma}
    For any feasible contract $\Omega(\Theta)=\{(q_k, p_k), \theta_k \in \Theta\}$, $q_k>q_j$ if and only if $\theta_k > \theta_j$, with the equality holds if and only if $\theta_k=\theta_j$.
\label{lemma:mono}
\end{lemma}
\begin{proof}
    For the sufficiency part, we need to show that if $\theta_k > \theta_j$, then $q_k>q_j$. Consider the following IC constraints, where $\theta_k, \theta_j \in \Theta^2, k \neq j$,
    \begin{align*}
        u_k(q_k, p_k) = \theta_k v(q_k) - p_k &\geq u_k(q_j, p_j) = \theta_k v(q_j) - p_j,\\
        u_j(q_j, p_j) = \theta_j v(q_j) - p_j &\geq u_j(q_k, p_k) = \theta_j v(q_k) - p_k.
    \end{align*} By adding them together, we can get
    \begin{equation*}
        \theta_k v(q_k) + \theta_j v(q_j) \geq \theta_k v(q_j) + \theta_j v(q_k).
    \end{equation*} Then, we have
        $v(q_k)(\theta_k-\theta_j) \geq v(q_j)(\theta_k-\theta_j).$ 
    Dividing both sides by $(\theta_k-\theta_j)$ gives us $v(q_k) > v(q_j)$. Since $v(\cdot)$ is a strictly increasing function of $q$, we have $q_k > q_j$.
    For the necessity part, we can show that if $q_k > q_j$, then $\theta_k > \theta_j$ through the same process, and a similar process is used for $q_k = q_j$ if and only if $\theta_k = \theta_j$.
\end{proof} 

\begin{proposition}
    For any feasible contract $\Omega(\Theta)=\{(q_k, p_k), \theta_k \in \Theta\}$, the QoS $q_k, k=1, \cdots, K$ satisfy the following monotonicity condition:
    \begin{equation*}
        0 \leq q_1 < \cdots < q_k < \cdots < q_K.
    \end{equation*}
\label{prop:r_mono}
\vspace{-5mm}
\end{proposition}
\begin{proof}
    The proof follows from Definition \ref{def:types} and Lemma \ref{lemma:mono}.
\end{proof}
Proposition \ref{prop:r_mono} is intuitive, as it implies that the user with a higher type (with a greater willingness to pay for better QoS, i.e., when QoS is more critical for the task it handles) will receive higher QoS. 

\begin{lemma}
    For any feasible contract $\Omega(\Theta)=\{(q_k, p_k), \theta_k \in \Theta\}$, the utility of each type of user must satisfy the following:
    \begin{equation*}
        0 \leq u_1(q_1, p_1) < \cdots < u_k(q_k, p_k) < \cdots < u_K(q_K, p_K).
    \end{equation*}
\label{lemma_u_mono}
\vspace{-5mm}
\end{lemma}
\begin{proof} 
    If $\theta_k>\theta_j$, we have
    \begin{equation*}
        \theta_k v(q_k) - p_k \geq \theta_k v(q_j) - p_j > \theta_j v(q_j) - p_j,
    \end{equation*} where the first inequality follows from the IC constraint while the second follows by $\theta_k>\theta_j$. Hence, $u_k(q_k, p_k) > u_j(q_j, p_j)$. The proof then follows from Definition \ref{def:types}.
\end{proof}


\begin{lemma}
    The IR constraints for problem \eqref{prob:opt_contract} will hold if the IR constraint for $\theta_1$ user is satisfied.
    \begin{equation}
        \theta_1 u_1(q_1, p_1) \geq 0.
        \label{eq:single_IR}
    \end{equation}
\label{lemma_reduce_IR}
\vspace{-5mm}
\end{lemma}
\begin{proof}
    The proof follows Lemma \ref{lemma_u_mono}.
\end{proof}
Lemma \ref{lemma_reduce_IR} indicates that if the utility of the lowest type satisfies the IR constraint, the entirety of the IR constraints set will hold, which reduces the total number of $K$ IR constraints into a single constraint, as in \eqref{eq:single_IR}.

\begin{lemma}
    The IC constraints for problem \eqref{prob:opt_contract} can be reduced to the following constraints:
    \begin{equation}
        \theta_{k} v(q_{k}) - p_{k} \geq \theta_{k} v(q_{k-1}) - p_{k-1}, \quad i = 2, \cdots, K.
        \label{eq:reduce_IC}
    \end{equation}
\label{lemma:reduce_IC}
\vspace{-5mm}
\end{lemma}
\begin{proof}
    The proof for reducing the number of IC constraints relies on the notions of local downward incentive constraints (LDICs) and local upward incentive constraints (LUICs). LDIC refers to the IC constraint between adjacent types $\theta_k$ and $\theta_{k-1}$, while LUIC refers to the IC constraint between types $\theta_k$ and $\theta_{k+1}$. Consider three types of users which follows $\theta_{k-2}<\theta_{k-1}<\theta_k$, which give us the following two LDICs:
    \begin{align*}
        \theta_{k} v(q_{k}) - p_{k} \geq \theta_{k} v(q_{k-1}) - p_{k-1},\\
        \theta_{k-1} v(q_{k-1}) - p_{k-1} \geq \theta_{k-1} v(q_{k-2}) - p_{k-2}.
    \end{align*} From the second inequality, we can obtain
    \begin{equation*}
        \theta_{k-1} (v(q_{k-1}) - v(q_{k-2}))\geq p_{k-1} - p_{k-2}.
    \end{equation*} Since $\theta_k > \theta_{k-1}$, and $q_{k-1}>q_{k-2}$ from Proposition \ref{prop:r_mono}, it follows that
    \begin{align*}
        \theta_{k} (v(q_{k-1}) - v(q_{k-2})) &\geq \theta_{k-1} (v(q_{k-1}) - v(q_{k-2}))\\ &\geq p_{k-1} - p_{k-2}.
    \end{align*} Therefore, we have 
    \begin{equation*}
         \theta_{k} v(q_{k-1}) - p_{k-1} \geq \theta_{k} v(q_{k-2}) - p_{k-2}.
    \end{equation*}
    Combining this with the first inequality gives
    \begin{equation*}
        \theta_{k} v(q_{k}) - p_{k} \geq \theta_{k} v(q_{k-1}) - p_{k-1} \geq \theta_{k} v(q_{k-2}) - p_{k-2}.
    \end{equation*} This implies that if the local downward incentive constraint (LDIC) holds for type $\theta_k$ user with respect to type $\theta_{k-1}$ user, then the downward incentive constraint (DIC) for type $\theta_k$ user with respect to type $\theta_{k-2}$ user will also hold. This reasoning can then extend for DICs down to type $\theta_1$. Hence, with the LDIC constraints for all $k=2, \cdots, K$, all the DICs hold. A similar process can be used to prove that with the LUIC constraints for all $k=1, \cdots, K-1$, all the UICs hold. Then, with monotonicity in Proposition \ref{prop:r_mono}, LDICs also imply that LUICs can be satisfied, which completes the proof.
\end{proof}

The problem \eqref{prob:opt_contract} can then be reduced to the following.
\begin{equation}
    \begin{aligned}
        \max_{\Omega(\Theta)} \ & \sum_{k=1}^{K}P(\theta_k)u_{SP}(q_k, p_k)\\
         \text{s.t.} \ & u_{1}(q_1, p_1) \geq 0, \ \text{(reduced IR)},\\
        & u_{k}(q_{k}, p_{k}) \geq u_{k}(q_{k-1}, p_{k-1}), k=2 \cdots K \ \text{(reduced IC)}, \\
        & 0 \leq q_1 < \cdots < q_k < \cdots < q_K \leq 1.
    \end{aligned}
\label{prob:reduce_contract}
\end{equation} Since the number of IR and IC constraints has been reduced, the problem \eqref{prob:reduce_contract} can now be solved using the Lagrangian multiplier method by relaxing the last constraint. This relaxation can serve as a projection function that ensures the feasibility of the solutions.

\section{Simulation Results}

We consider the design of a pricing menu for cloud-based agentic AI services applied to a cybersecurity task such as system log analysis. An illustrative configuration with varying QoS levels is presented in Table \ref{tab:ex1}. Typically, $D^{\text{in}}_k=20, 100$ represent small and large input log files, respectively. Similarly,  $D^{\text{out}}_k=20, 100$ correspond to shorter outputs (e.g., binary results with brief explanations) and longer outputs (e.g., detailed reasoning and analysis with tables or figures). The LLM model-related parameters $\beta_k, n^{\text{layer}}_k, n^{\text{ctx}}_k, n^{\text{attn}}_k$ are selected from GPT-2 Small ($k=1, 2, 3$), Phi-2 ($k=4, 5, 6$), Gemma 7B ($k=7, 8$), respectively. Due to the absence of publicly available user satisfaction data for these models in this task domain, we assume satisfaction increases with larger input/output sizes and more powerful models. The associated computational capacities $\gamma_k$ are based on NVIDIA GPUs (T4, L4, A100, A10), with performance ranging from 8,100 to 31,200 GFLOPS.

\begin{table}[h!]
\centering
\caption{Example LLM Service Configurations}
 \begin{tabular}{|c c c c c c c c|} 
 \hline 
  $k$ &  $D^{\text{in}}_k$ &  $D^{\text{out}}_k$  &   $\beta_k$ &  $n^{\text{layer}}_k, n^{\text{ctx}}_k, n^{\text{attn}}_k$ & $A_k$ & $\gamma_k$ & $q_k$ \\
 \hline\hline 
1 & 20 & 20 & 0.12 & 12, 1024, 12 & 0.1 & 8100 &  .531\\
2 & 100 & 20 & 0.12 & 12, 1024, 12 & 0.22 & 8100 & .555 \\
3 & 100 & 100 & 0.12 & 12, 1024, 12 & 0.35 & 15800 & .584 \\
4 & 20 & 20 & 2.7 &  32, 2048, 32 & 0.35 & 15800 & .655 \\
5 & 100 & 20 & 2.7 &  32, 2048, 32 & 0.5 & 19500 & .691 \\
6 & 100 & 100 & 2.7 &  32, 2048, 32 & 0.65 & 19500 & .728 \\
7 & 100 & 20 & 7 & 28, 8192, 16 &  0.75 & 31200  & .814 \\
8 & 100 & 100 & 7 &  28, 8192, 16 & 0.9 & 31200  & .848 \\
[0.5ex] 
 \hline
 \multicolumn{8}{c}{$\beta_k, n^{\text{layer}}_k, n^{\text{ctx}}_k, n^{\text{attn}}_k$ from GPT-2 Small, Phi-2, Gemma 7B.}\\
 \multicolumn{8}{c}{$\gamma_k$ from NVIDIA GPU T4, L4, A100, A10.}
 \label{tab:ex1}
 \end{tabular}
 \vspace{-3mm}
\end{table}

\begin{figure}
    \centering
    \includegraphics[width=0.7\linewidth]{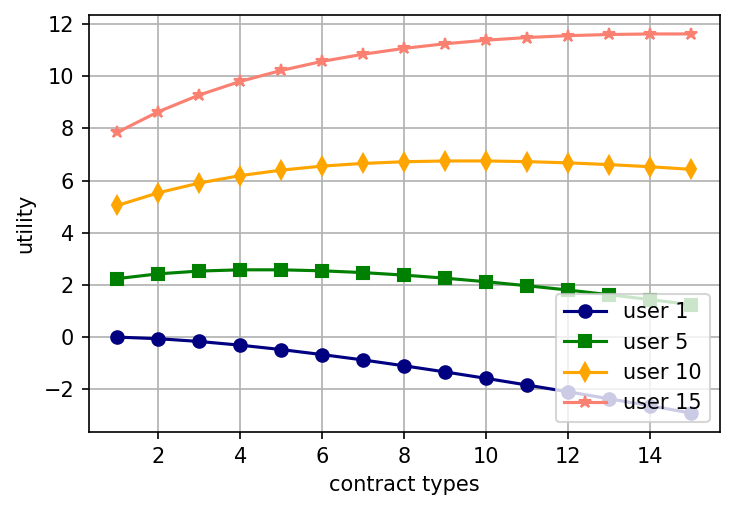}
    \caption{User utilities for types 1,5,10,15 under different contract options. The results show that each user achieves the highest utility by selecting the contract designed for their type.}
    \label{fig:types_utility}
\vspace{-4mm}
\end{figure}

Consider a transmission rate of $r=20$ Mbps between the user and the cloud, a tokenization/detokenization time of $0.5$ ms per KB, and the functions $g_t(D_k^{\text{in}})=4D_k^{\text{in}}, g_{dt}(D_k^{\text{out}})=4D_k^{\text{out}}$, along with a task-specific parameter for preference $\delta=0.5$, the QoS corresponding to each configuration is computed and presented in the final column.
We then use the resulted $q_k$ and the corresponding computational cost value $C(q_k)$ from \eqref{eq:c_q} (without considering $c^{\text{l}}(q_k)$ here) to fit the relationship curve between $q$ and $C(q)$.
The results for user types $K=15$ with $\theta_k=k$ are shown in Figure \ref{fig:types_utility}. As shown, each user achieves the highest and non-negative utility by selecting the contract corresponding to their own type. This behavior confirms that the designed contracts satisfy the individual rationality and incentive compatibility constraints, as defined in Definitions \ref{def:cons_IR} and \ref{def:cons_IC}, respectively. 

To examine how potential liability costs influence the contract design, for each configuration in Table \ref{tab:ex1}, we incorporate liability cost for the service provider. We assume that larger and more sophisticated models are less prone to hallucination and misalignment. Accordingly, we assign liability costs as $c^{\text{l}}(q_k)=0.7, 0.5, 0.2$ for $q_k$ parameters derived from GPT-2 Small, Phi-2, and Gemma 7B, respectively. Note that this may not always be the case since larger models often express hallucinated content more fluently and confidently, which can make them more misleading when errors occur.
We then use the resulting $q_k$ and the corresponding cost value $C(q_k)$ to fit the relationship curve between $q$ and $C(q)$.
Besides, we also consider the case where there is no information asymmetry (first-best). In this case, SP knows the exact user type $\theta_k$.

\begin{figure}
    \centering
    \includegraphics[width=0.72\linewidth]{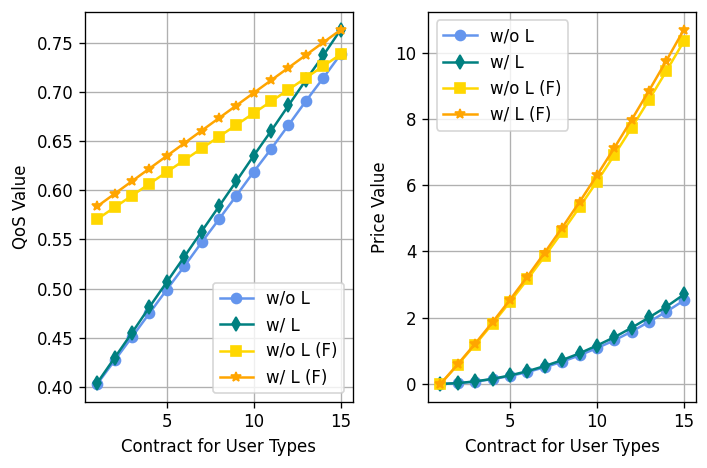}
    \caption{The QoS levels and corresponding prices of contracts for different user types. Here, L indicates the liability cost, and (F) represents first-best results.}
    \label{fig:qpc}
\vspace{-4mm}
\end{figure}

\begin{figure}
    \centering
    \includegraphics[width=0.72\linewidth]{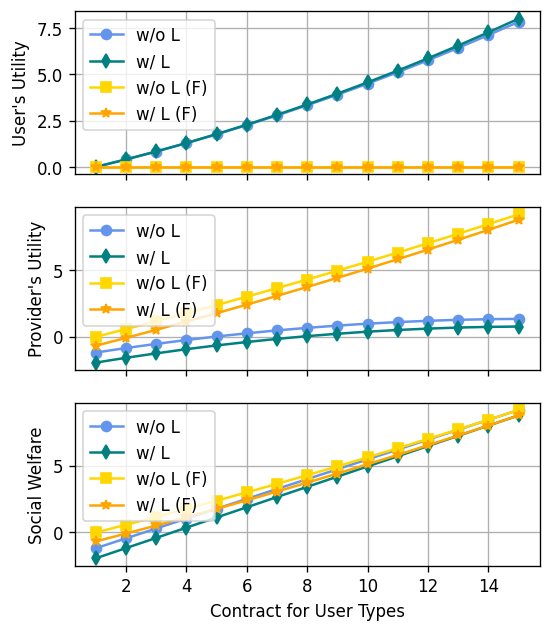}
    \caption{The user's and provider's utilities as well as the social welfare (sum of the user's and provider's utilities).}
    \label{fig:sw}
\vspace{-4mm}
\end{figure}

Figure \ref{fig:qpc} shows the resulting QoS levels and corresponding prices for each user type, with and without incorporating liability costs. Introducing liability costs leads to a slight increase in both QoS levels and prices across user types, implying that the SP may have the incentive to provide better services. Figure \ref{fig:sw} illustrates the utilities for the users, the SP, and overall social welfare. Under complete information, the users’ utilities become zero while the SP’s utility increases. However, under information asymmetry, the SP’s expected utility, as defined in \eqref{eq:U_sp}, shifts from positive to negative when liability costs are introduced. This highlights the importance of careful liability cost design from a third-party or regulatory perspective. If the liability cost is set too low, the SP may lack sufficient incentive to mitigate hallucinations or output risks. Conversely, if it is excessively high, the SP may be disincentivized from continuing to offer the service.

\section{Conclusion}
In this work, we propose PACT, a holistic contract-theoretic framework for pricing task-specific cloud-based agentic AI services based on Quality of Service (QoS) across domains such as cybersecurity, healthcare, and business operations. The framework models QoS as a combination of objective metrics, such as response time, and subjective metrics, such as user satisfaction, to address the challenge of task-dependent and multi-dimensional service evaluation from the user’s perspective. To reflect future regulatory and ethical risks in high-stakes domains, we incorporate liability costs into the service provider’s cost structure. The proposed contract-based pricing mechanism ensures user participation while discouraging overuse and resource inefficiency by satisfying incentive compatibility that aligns service quality with user needs.
Future research directions include designing auditing mechanisms for verifying agentic AI service outcomes and developing principled approaches to setting liability costs that balance the service provider’s incentive to continue offering services with the need to reduce hallucinations and misaligned outputs.

\bibliography{reference}
\bibliographystyle{IEEEtran}

\end{document}